\documentclass{PoS}
\usepackage{graphicx}
\usepackage[caption=false]{subfig}
\usepackage{bm}
\usepackage{amssymb}
\usepackage{amsmath}
\usepackage{booktabs}
\usepackage[utf8]{inputenc}
\usepackage{microtype}
\usepackage{siunitx}
\usepackage[capitalize]{cleveref}

\newcommand{\nusquids}{\texttt{nuSQuIDS}}
\newcommand{\emcee}{\texttt{emcee}}
\newcommand\barparen[1]{\overset{\scriptscriptstyle(-)}{#1}}

\title{Measurement of the high-energy all-flavor neutrino-nucleon cross section with IceCube}

\ShortTitle{Neutrino-nucleon cross section with IceCube}

\author{
The IceCube Collaboration\footnote{For collaboration list, see PoS(ICRC2019) 1177.}\\
{\itshape \href{http://icecube.wisc.edu/collaboration/authors/icrc19_icecube}{http://icecube.wisc.edu/collaboration/authors/icrc19\_icecube}}\\
E-mail: \email{tyuan@icecube.wisc.edu}
}

\abstract{
The flux of high-energy neutrinos passing through the Earth is attenuated due to their interactions with matter. Their transmission probability is modulated by the neutrino interaction cross section and affects the arrival flux at the IceCube Neutrino Observatory, a cubic-kilometer neutrino detector embedded in the South Pole ice sheet. We present a measurement of the neutrino-nucleon cross section between 60~TeV--10~PeV using the high-energy starting events (HESE) sample from IceCube with 7.5 years of data.\\

\vspace{4mm}
{\bfseries Corresponding authors:}
\speaker{Tianlu Yuan}$^{1}$\\
{$^{1}$ \itshape Dept. of Physics and Wisconsin IceCube Particle Astrophysics Center, University of Wisconsin, Madison, WI 53706, USA}

}

\FullConference{36th International Cosmic Ray Conference -ICRC2019-\\
		July 24th - August 1st, 2019\\
		Madison, WI, U.S.A.}

\begin{document}

\section{Introduction}\label{sec:intro}

Neutrinos above TeV energies that traverse through the Earth may interact before exiting~\cite{Gandhi:1995tf}. At these energies neutrino-nucleon interactions proceed via deep-inelastic scattering (DIS), whereby the neutrino interacts with the constituent quarks within the nucleon. The DIS cross sections can be derived from parton distribution functions (PDF) which are in turn constrained experimentally~\cite{CooperSarkar:2011pa} or by using a color dipole model of the nucleon and assuming that cross-sections increase at high energies as $\ln^2 s$~\cite{Arguelles:2015wba}. At energies above a PeV, more exotic effects beyond the Standard Model have been proposed that predict a neutrino cross section of up to \SI{e-27}{\cm^2} at $E_\nu > \SI{e19}{eV}$~\cite{Jain:2000pu}. Thus far, measurements of the high-energy neutrino cross section have been performed using data from the IceCube Neutrino Observatory. One proposed experiment, the ForwArd Search ExpeRiment at the LHC (FASER), plans to measure the neutrino cross section at TeV energies~\cite{Ariga:2019ufm}.

The IceCube Neutrino Observatory is a cubic-kilometer neutrino detector installed in the ice at the geographic South Pole~\cite{Aartsen:2016nxy}, between depths of \SI{1450}{\m} and \SI{2450}{\m}, completed in 2010. Reconstruction of the direction, energy and flavor of the neutrinos relies on the optical detection of Cherenkov radiation emitted by charged particles produced in the interactions of neutrinos in the surrounding ice or the nearby bedrock. As the transmission probability through the Earth is dependent on the neutrino cross section, a change in the cross section affects the arrival flux of neutrinos at IceCube as a function of the energy and zenith angle. Recently, IceCube performed the first measurement of the high-energy neutrino-nucleon cross section using a sample of upgoing muon neutrinos~\cite{Aartsen:2017kpd}. In this paper, we present a measurement of the neutrino-nucleon cross section using the high-energy starting events (HESE) sample with 7.5 years of data~\cite{Schneider:2019icrc_hese}. By using events that start in the detector, the measurement is sensitive to both the northern and southern skies, as well as all three flavors of neutrinos, unlike~\cite{Bustamante:2017xuy} which used only a single class of events in the six-year HESE sample.

\section{Analysis method}\label{sec:method}

\begin{figure}[htb]
    \centering
    \subfloat{
        \includegraphics[width=0.46\textwidth]{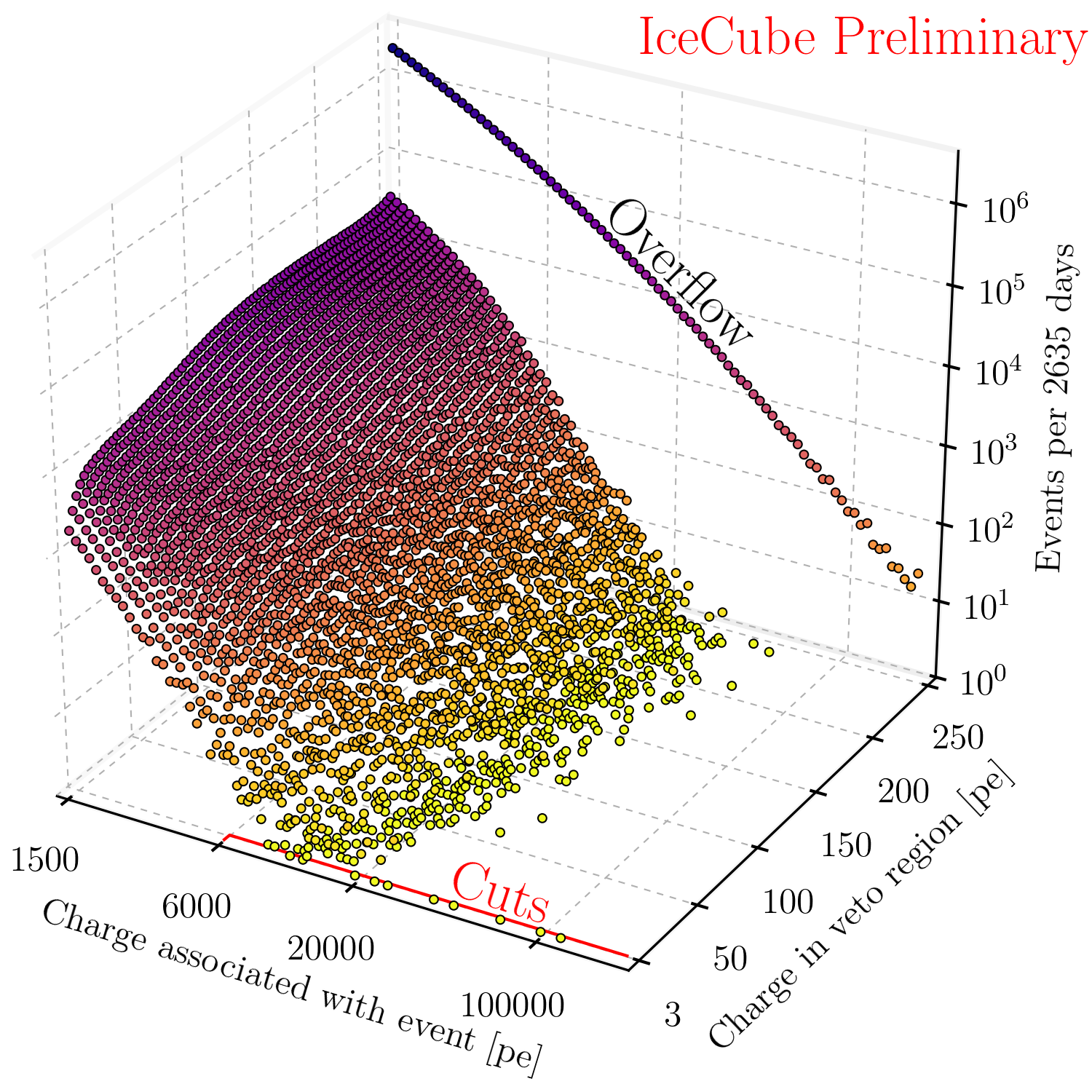}
    }
    \subfloat{
        \includegraphics[width=0.52\textwidth]{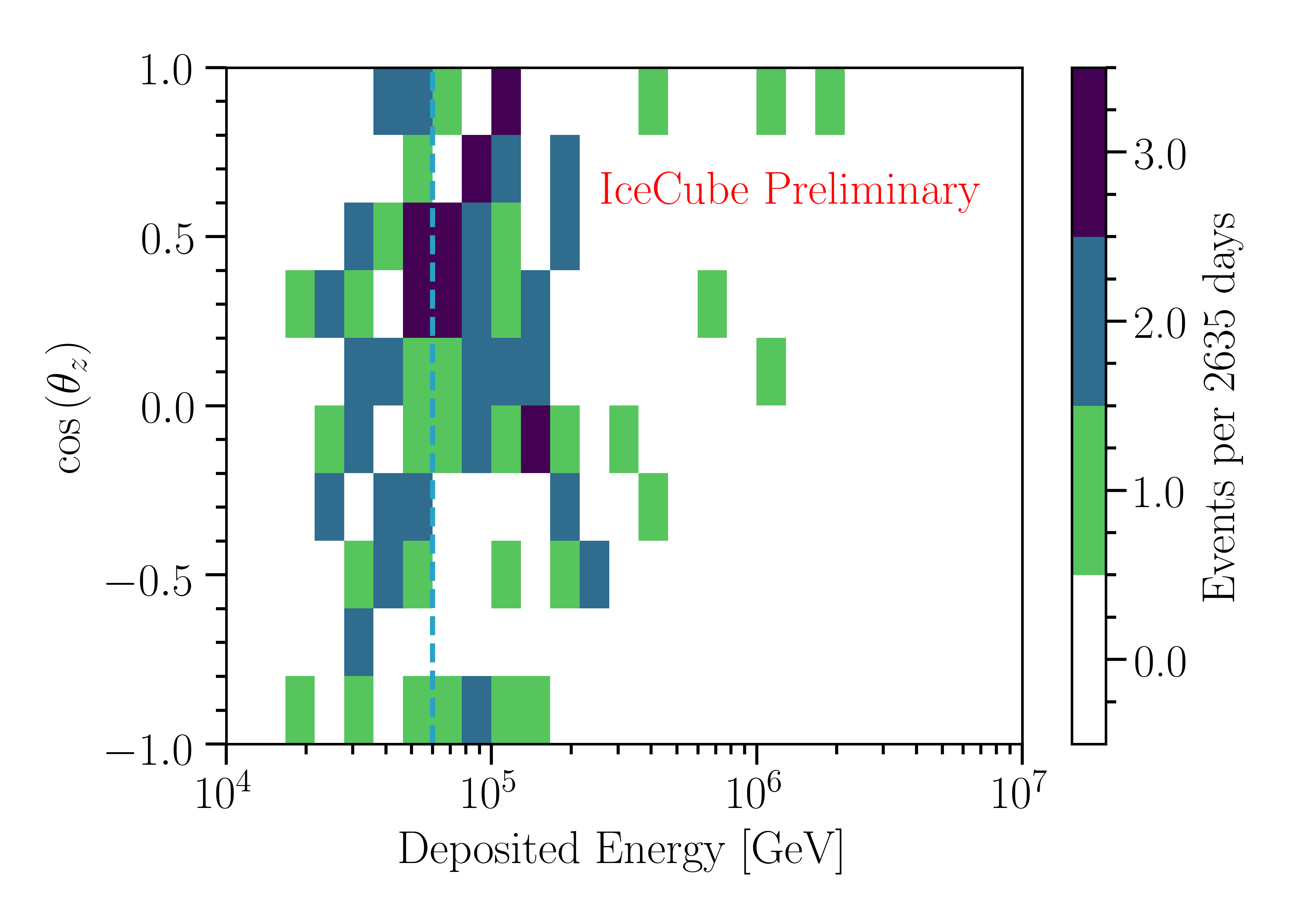}
    }
    \caption{The left panel shows the distribution of background and signal events for the HESE selection in terms of charge in the veto region and total event charge. The charge cuts are shown by the solid red lines where the signal is defined to be in the region with greater than \SI{6000}{photoelectrons} for charge associated with the event and less than \SI{3}{photoelectrons} in the veto region. The color scale corresponds to the z-axis, which is the number of events in that bin per 2635 days. The right panel shows the distribution of signal events as a function of their inferred deposited energy and cosine of the reconstructed zenith angle. The dashed line indicates the low-energy threshold of 60 TeV.}
    \label{fig:qtotveto}
\end{figure}

Several improvements have been incorporated into the HESE-7.5 analysis chain, and are used in this measurement. These include better detector modeling, a three-topology classifier that corresponds to the three neutrino flavors~\cite{Usner:2018qry}, improved atmospheric neutrino background calculation~\cite{Arguelles:2018awr}, and a new likelihood treatment that accounts for statistical uncertainties~\cite{Arguelles:2019izp}. The selection cuts have remained unchanged and require the total charge associated with the event to be above \SI{6000}{photoelectrons} with the charge in the outer layer of the detector (veto region) to be below \SI{3}{photoelectrons}. This rejects almost all of the atmospheric muon background, as well as a fraction of atmospheric neutrinos from the southern sky that are accompanied by muons, as shown in the left panel of \cref{fig:qtotveto}.  There are a total of 102 events that pass the charge cuts. A histogram of their deposited energy and reconstructed cosine zenith angle is shown in the right panel of \cref{fig:qtotveto}. For this analysis, only the 60 events with reconstructed energy above \SI{60}{TeV} are used. A forward-folded likelihood is constructed using deposited energy and $\cos(\theta_z)$ distributions for tracks and cascades separately. For the two double cascades above \SI{60}{TeV} the likelihood is constructed using a distribution of the cascade-length separation and deposited energy.

\begin{figure*}[htp!]
\centering
    \subfloat{
        \includegraphics[width=0.49\textwidth]{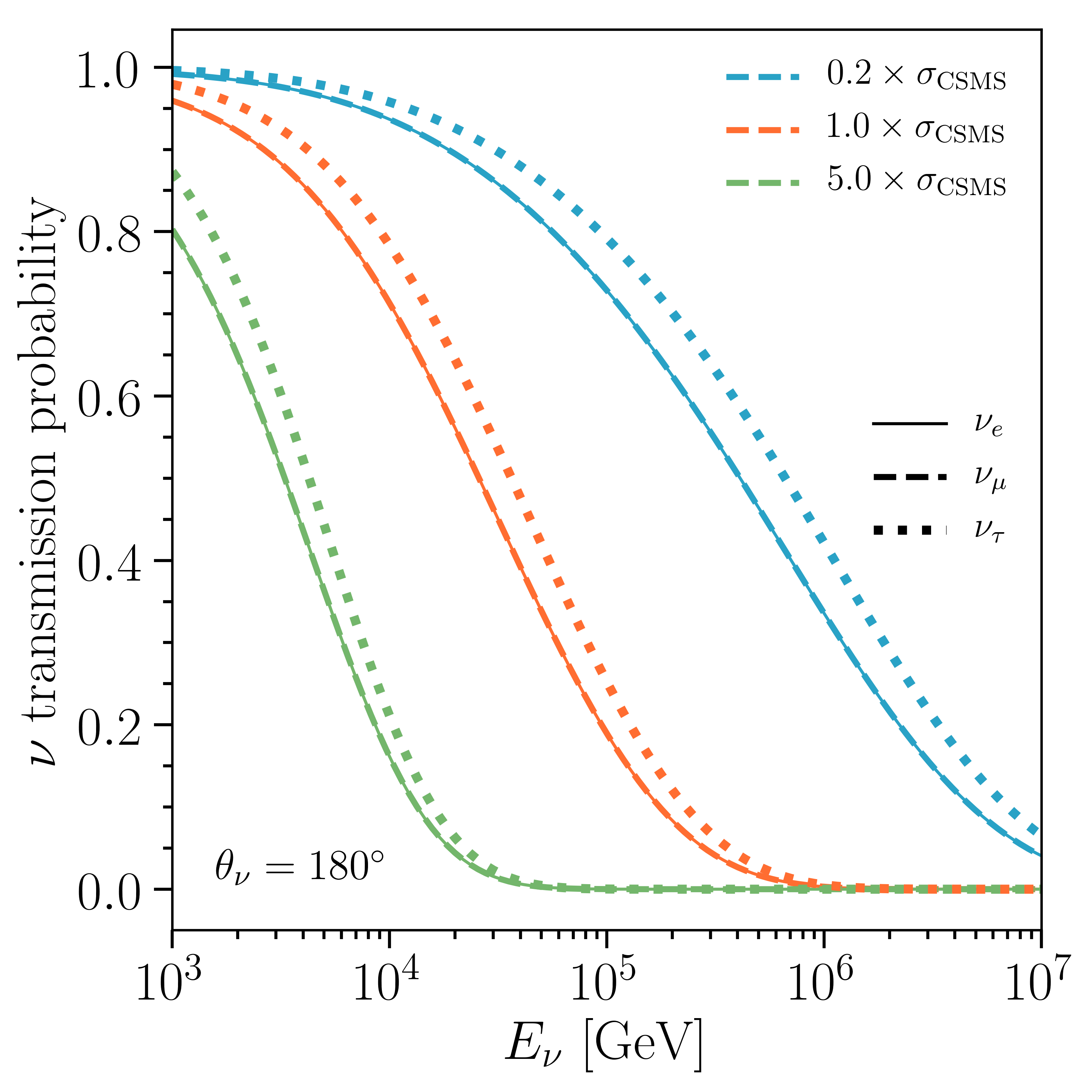}
    }
    \subfloat{
        \includegraphics[width=0.49\textwidth]{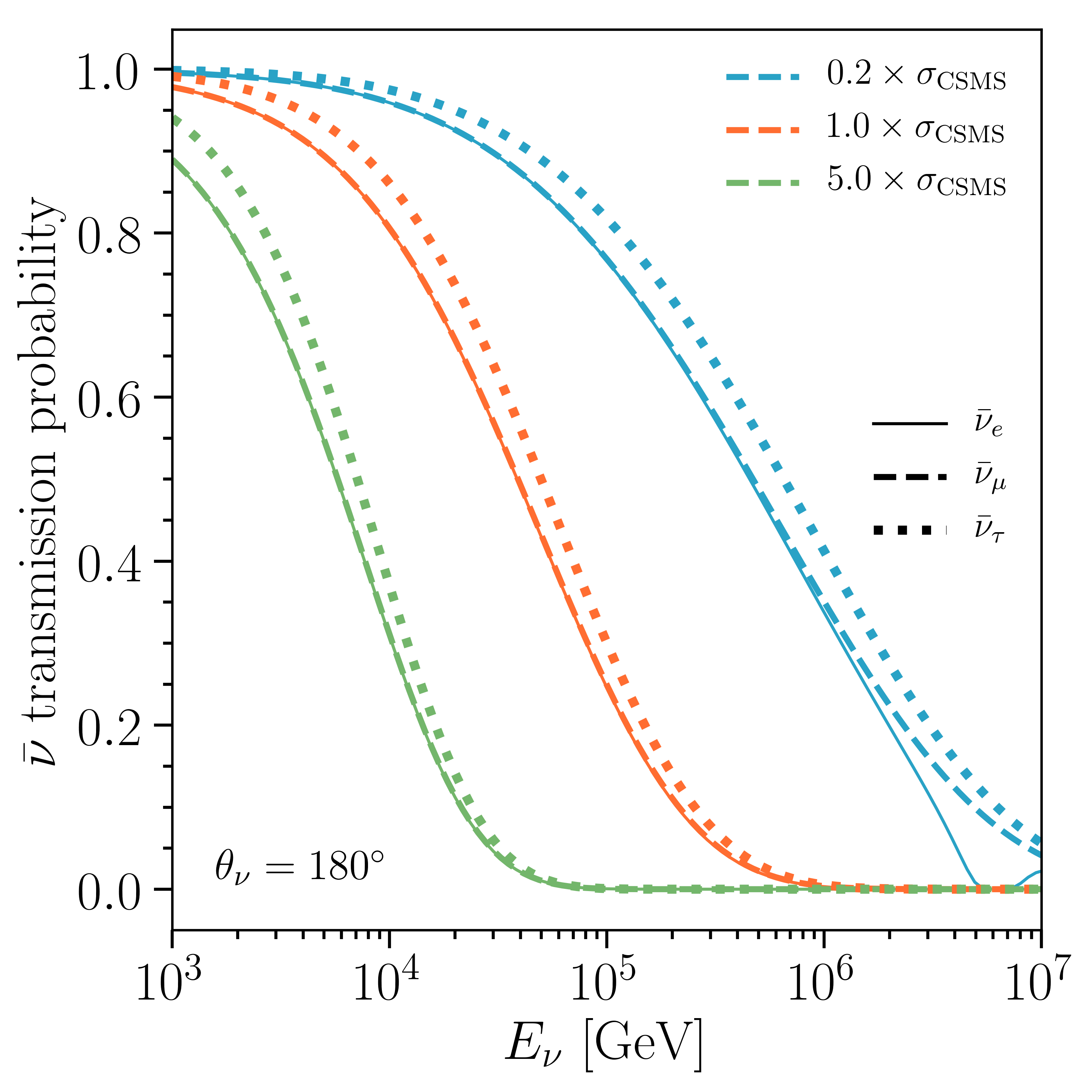}
    }\\
    \subfloat{
        \includegraphics[width=0.49\textwidth]{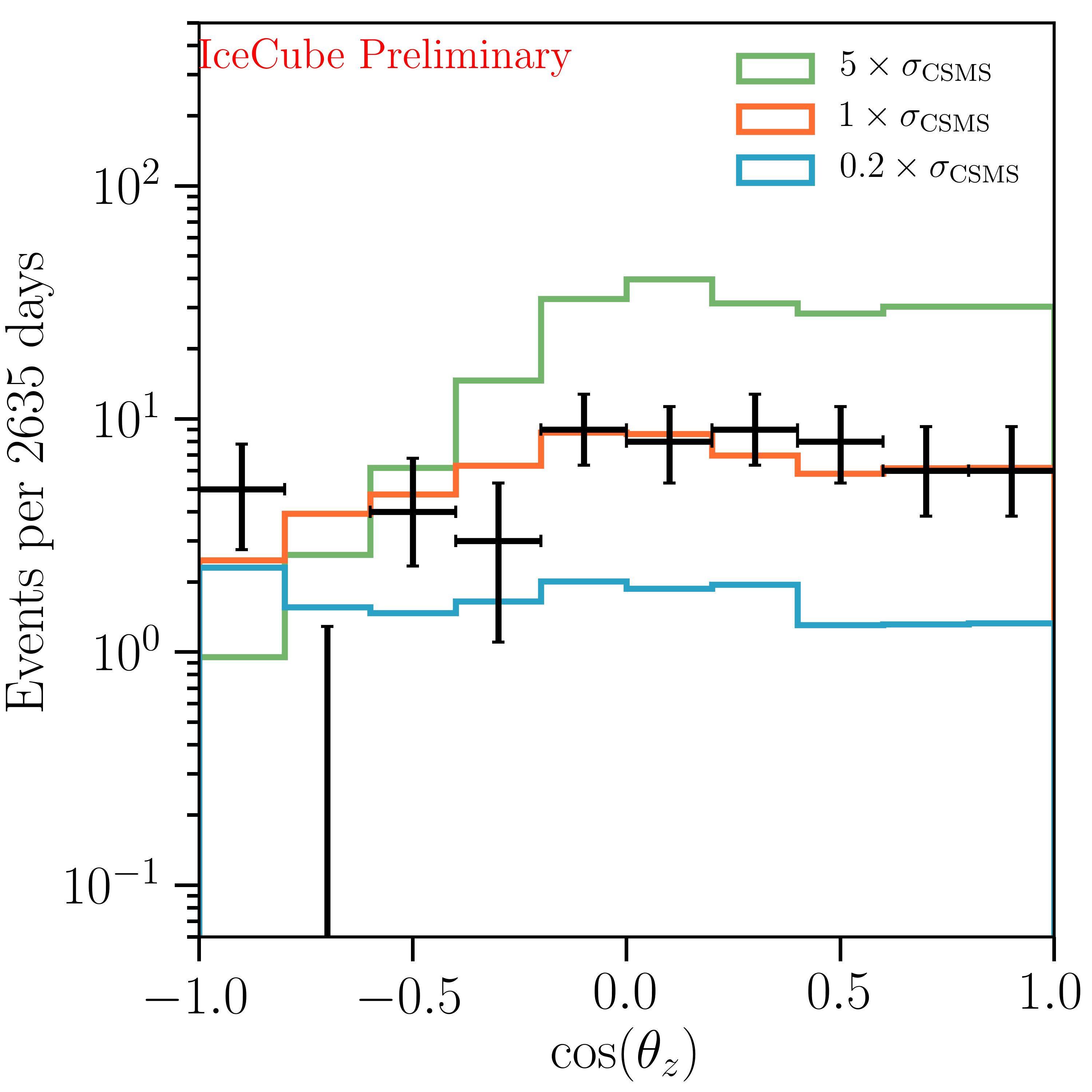}
    }

	\protect\caption{Upgoing neutrino (top left) and antineutrino (top right) transmission probabilities as a function of $E_\nu$ for three realizations of the cross section. The flux at the surface is assumed to have a spectral index of $\gamma=2$. The scaling is applied to the cross section given in~\cite{CooperSarkar:2011pa}. The bottom panel shows the effect of three different DIS cross sections on the marginalized one-dimensional distribution in reconstructed $\cos(\theta_z)$. The data is indicated by black crosses and MC expectations for a fixed single-power-law flux shown in the solid lines. The expected distribution assuming the nominal CSMS cross section is shown in orange.}
    \label{fig:attenuation1d}
\end{figure*}

Neutrino (top left) and antineutrino (top right) transmission probabilities are shown in \cref{fig:attenuation1d} for three different variations of the DIS cross section given in~\cite{CooperSarkar:2011pa} (CSMS). They are plotted for each flavor individually as a function of the neutrino energy, $E_\nu$, at $\theta_\nu=180^\circ$, assuming an initial surface flux with spectral index of $\gamma=2$. As the cross section is decreased, the transmission probability increases since neutrinos are less likely to interact on their way through the Earth. On the other hand, a higher cross section implies a higher chance of interaction and the transmission probability decreases. The reason there is a slight flavor-dependence is due to the fact that charged current (CC) $\barparen{\nu}_e$ and $\barparen{\nu}_\mu$ interactions produce charged particles that rapidly lose energy in matter, while a CC $\barparen{\nu}_\tau$ interaction produces a tau lepton, which can immediately decay to a slightly lower energy $\barparen{\nu}_\tau$. Neutral current (NC) interactions are also non-destructive, producing a secondary neutrino at a slightly lower energy than the parent~\cite{Vincent:2017svp}. Furthermore, there is a dip in the $\bar{\nu}_e$ transmission probability due to the Glashow resonance (GR)~\cite{Glashow:1960zz}.

In this analysis, the neutrino-nucleon cross section is measured as a function of energy by dividing the CSMS cross section into four bins: \SIrange{60}{100}{TeV}, \SIrange{100}{200}{TeV}, \SIrange{200}{500}{TeV}, and \SIrange{500}{10000}{TeV}. The overall normalization of the cross section in each bin is allowed to float with four scaling parameters $\bm{x}=(x_0, x_1, x_2, x_3)$, where the index goes from the lowest energy bin to the highest energy bin. We further assume that the ratio of the CC to NC cross section is fixed and that there is no additional flavor dependence. Thus, $\bm{x}$ is applied identically across all flavors and interaction channels on the CSMS prediction. In order to model the effect of varying the cross section on the arrival flux, we used \nusquids~\cite{Delgado:2014kpa}. This allows us to account properly for destructive CC interactions as well as for secondaries from NC interactions and tau-regeneration. The Earth density is set to the preliminary reference Earth model (PREM)~\cite{Dziewonski:1981xy} and the GR cross section is kept fixed to the Standard Model prediction. We also include the nuisance parameters given in \cref{tab:nuisances}, for a single-power-law astrophysical flux, pion and kaon induced atmospheric neutrino flux by Honda et~al., and BERSS prompt atmospheric neutrino flux~\cite{Honda:2006qj,Bhattacharya:2015jpa}.
%
\begin{table}[thb]
\centering
\begin{tabular}{l|rr}
Parameter & Constraint/Prior  & Range\\ 
\hline
\multicolumn{1}{l|}{\textbf{Astrophysical neutrino flux:}} & & \\
$\Phi_\texttt{astro}$ & - & $[0,\infty)$ \\
$\gamma_\texttt{astro}$ & $2.0\pm1.0$ &  $(-\infty,\infty)$ \\
&&\\
\multicolumn{1}{l|}{\textbf{Atmospheric neutrino flux:}} & &\\
$\Phi_\texttt{conv}$ & $1.0\pm0.4$ & $[0, \infty)$ \\
$\Phi_\texttt{prompt}$ & $1.0\pm3.0$ & $[0, \infty)$ \\
$\pi/K$ & $1.0\pm0.1$ & $(-\infty, \infty)$\\
${2\nu/\left(\nu+\bar{\nu}\right)}_\texttt{atmo}$ & $1.0\pm0.1$ & $[0,2]$ \\
&&\\
\multicolumn{1}{l|}{\textbf{Cosmic ray flux:}} & &\\
$\Delta\gamma_\texttt{CR}$ & $-0.05\pm 0.05$ & $(-\infty,\infty)$ \\
$\Phi_\mu$ & $1.0\pm 0.5$ & $[0,\infty)$ \\
\end{tabular}
\caption{Central values and uncertainties on the nuisance parameters included in the fit. Truncated Gaussians are set to zero for all negative parameter values.}
\label{tab:nuisances}
\end{table}

As $\bm{x}$ is varied, Monte-Carlo (MC) events are reweighted by the ratio of $x_i \Phi(E_\nu, \theta_\nu, \bm{x})/\Phi(E_\nu, \theta_\nu, \bm{1})$, where $\Phi$ is the arrival flux as calculated by \nusquids{}, $E_\nu$ is the true neutrino energy, $\theta_\nu$ the true neutrino zenith angle, and $x_i$ the scaling factor for the bin that covers $E_\nu$. The arrival flux is dependent on $\bm{x}$, while the linear factor $x_i$ is due to the increased probability of interaction at the detector. The MC provides a mapping from the true physics space to reconstructed quantities, and allows us to construct a likelihood using the reconstructed zenith and energy distributions for tracks and cascades, and reconstructed energy and cascade length separation distribution for double-cascades~\cite{Schneider:2019icrc_hese}. This likelihood can then be maximized (frequentist) or marginalized (Bayesian) to obtain the set of scalings that best describe the data, $\bm{\hat{x}}$. A likelihood scan over four dimensions was performed to obtain the frequentist confidence regions assuming Wilks' theorem. A MCMC sampler, \emcee~\cite{ForemanMackey:2012ig}, was used to obtain the Bayesian credible regions.

The effect of changing the overall cross section on the expected event rate in $\cos(\theta_z)$ is shown in the bottom panel of \cref{fig:attenuation1d}. Predictions from two alternative cross sections are shown along with the nominal CSMS expectations (orange), all assuming the best-fit, single-power-law flux from~\cite{Schneider:2019icrc_hese}. In the southern sky, $\cos(\theta_z) > 0$, the Earth absorption is negligible so the effect of rescaling the cross section is linear. In the northern sky, $\cos(\theta_z) < 0$, the strength of Earth absorption is dependent on the cross section and the expected number of events is seen to fall off towards $\cos(\theta_z)=-1$ for the nominal and $5\times \sigma_{\rm CSMS}$ (green) cases. This decreased expectation in the northern sky can also be seen in the right panel of \cref{fig:qtotveto}, from which a relatively fewer number of events arrive as compared to the southern sky.

\section{Results}\label{sec:results}

\begin{figure*}[htp]
\centering
    \subfloat{
        \includegraphics[width=0.49\textwidth]{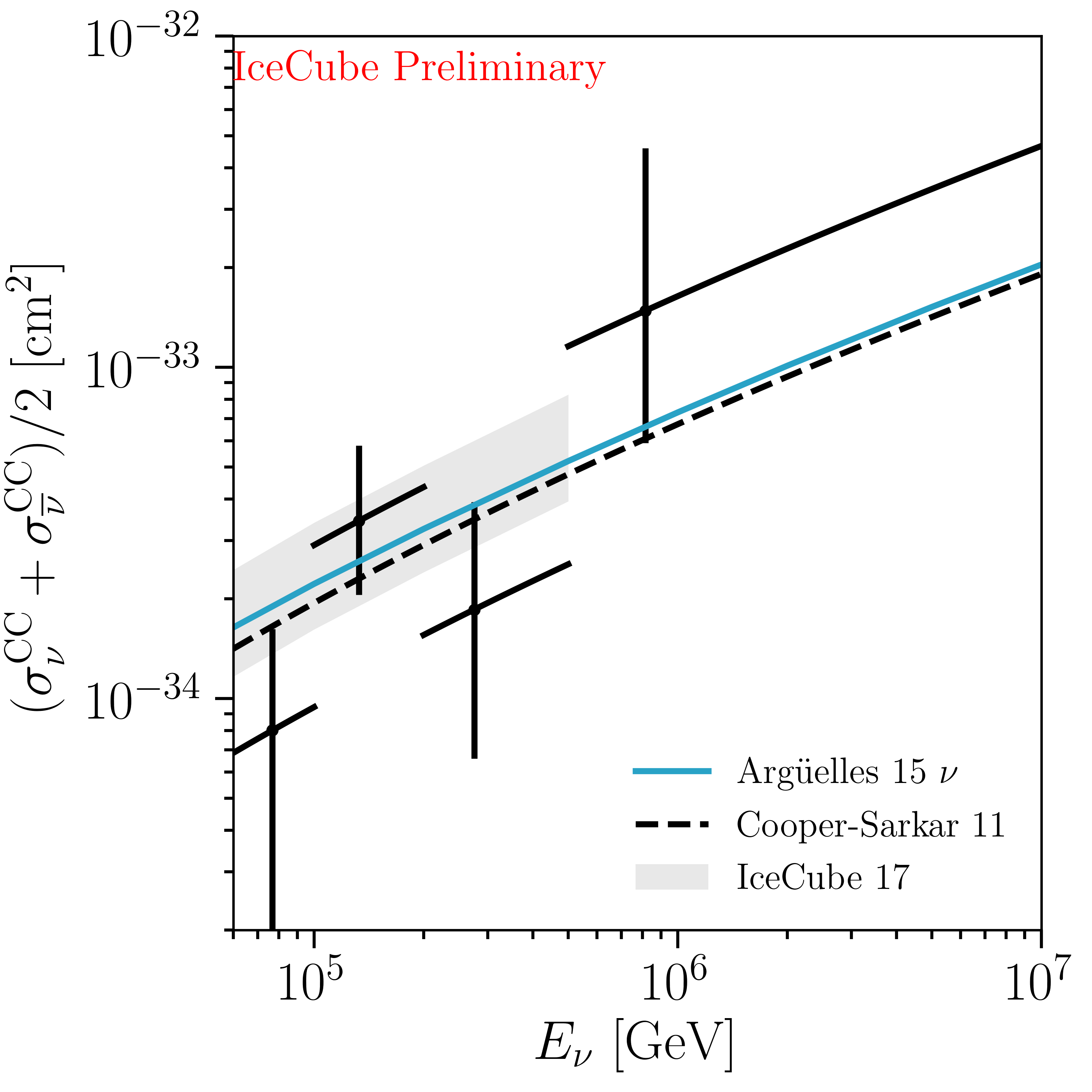}
    }
    \subfloat{
        \includegraphics[width=0.49\textwidth]{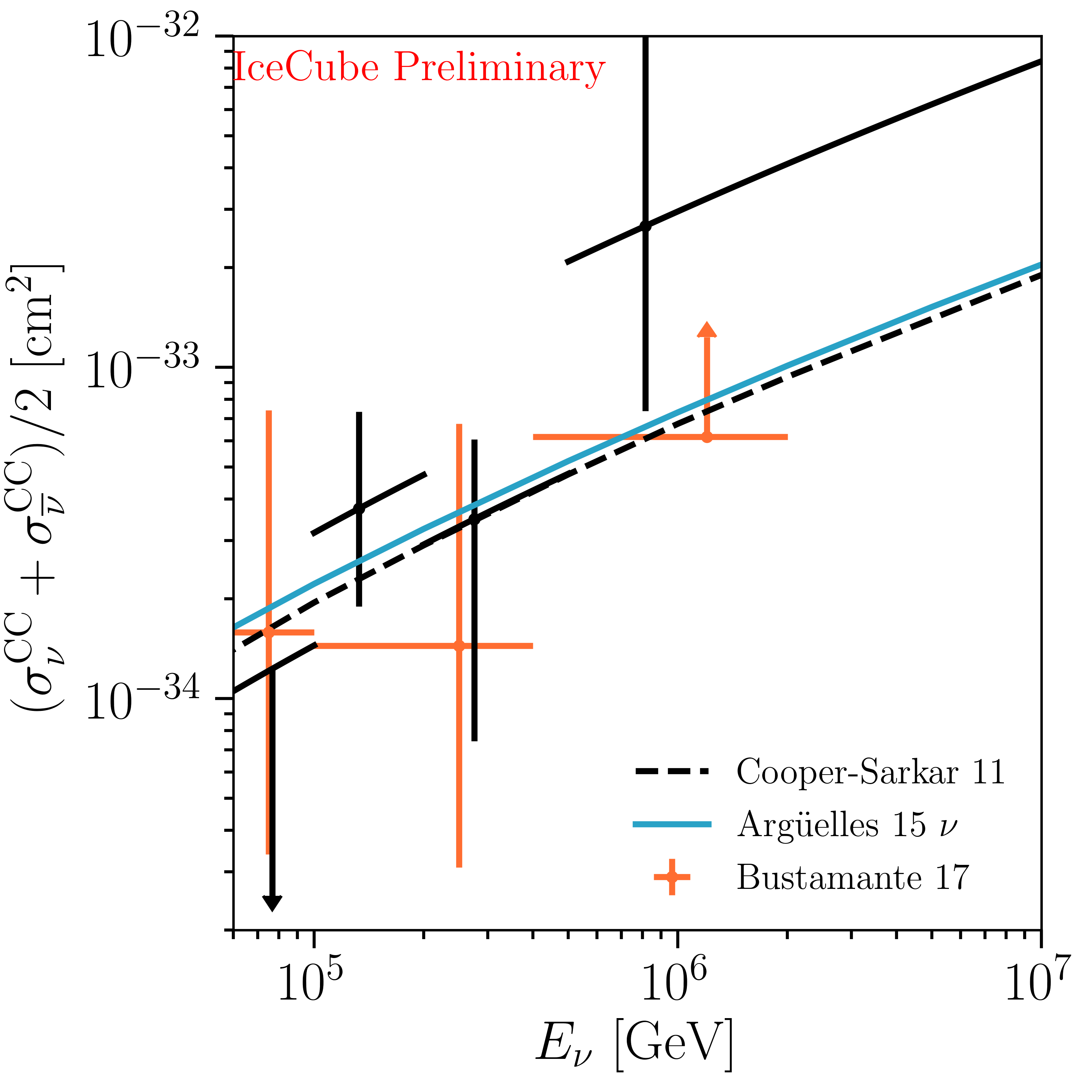}
    }

	\protect\caption{Measured neutrino DIS cross section using the HESE sample with 7.5 years of livetime are shown as black error bars. The left panel shows the frequentist 68.3\% confidence intervals assuming Wilks' theorem. The right panel shows the Bayesian 68.3\% credible intervals. For comparison, the results from~\cite{Aartsen:2017kpd} and~\cite{Bustamante:2017xuy} are included in the left and right panels, respectively. The prediction from~\cite{Arguelles:2015wba} is shown as the blue, solid line, and the prediction from~\cite{CooperSarkar:2011pa} is shown as the black, dashed line}
    \label{fig:results}
\end{figure*}

\Cref{fig:results} shows the frequentist 68.3\% confidence interval (left panel) and Bayesian 68.3\% credible interval (right panel) on the CC cross section obtained using the HESE 7.5 sample. For comparison of frequentist results, the measurement from~\cite{Aartsen:2017kpd} (gray region) is included in the left panel. For comparison of Bayesian results, the measurement from from~\cite{Bustamante:2017xuy} (orange error bars) is included in the right panel. The prediction from~\cite{Arguelles:2015wba} is shown as the solid, blue line and the CSMS cross section as the dashed, black line. As the ratio of CC to NC cross section is assumed to be fixed, the NC cross section is identical relative to the CSMS predictions and so is not shown here.

In both the frequentist and Bayesian results, the lowest-energy bin prefers a lower cross section while the highest-energy bin prefers a higher cross section than the Standard Model predictions. However, as the uncertainties are large, none of the bins are in significant tension with the models.

\section{Summary}\label{sec:summary}

In this proceeding, we have presented a measurement of the neutrino-nucleon cross section above 60~TeV using a sample of high-energy, starting events detected by IceCube. The measurement relies on the Earth as a neutrino attenuator, one that is dependent on the neutrino interaction rate and hence its cross section. The HESE sample used for this analysis spans 7.5 years of livetime with many improvements incorporated into the analysis chain. As the sample is of starting events, events from both the norther and southern sky are included, and with the new three-topology classifier, the likelihood utilizes flavor information as much as possible.

The results are obtained in both frequentist and Bayesian statistical frameworks. This allows for direct comparison to the two previously published measurements, with which the results here are consistent. Both frequentist and Bayesian results are consistent with Standard Model calculations. Though the current uncertainties are large, it will be possible to constrain the cross section to better precision with future, planned detector upgrades.

\bibliographystyle{ICRC}
\bibliography{references}

%

\end{document}